\documentclass[a4paper, 12pt]{article}

\usepackage{amsmath}

\begin{document}
\renewcommand{\theequation}{\thesection .\arabic{equation}}

\newcommand{\sign}{\operatorname{sign}}
\newcommand{\Ci}{\operatorname{Ci}}
\newcommand{\tr}{\operatorname{tr}}

\newcommand{\beq}{\begin{equation}}
\newcommand{\eeq}{\end{equation}}
\newcommand{\beqn}{\begin{eqnarray}}
\newcommand{\eeqn}{\end{eqnarray}}

\newcommand{\slp}{\raise.15ex\hbox{$/$}\kern-.57em\hbox{$ \partial $}}
\newcommand{\lnA}{\raise.15ex\hbox{$/$}\kern-.57em\hbox{$A$}}
\newcommand{\unmedio}{{\scriptstyle\frac{1}{2}}}
\newcommand{\uncuarto}{{\scriptstyle\frac{1}{4}}}

\newcommand{\trial}{_{\text{trial}}}
\newcommand{\true}{_{\text{true}}}
\newcommand{\const}{\text{const}}

\newcommand{\intp}{\int\frac{d^2p}{(2\pi)^2}\,}
\newcommand{\intx}{\int d^2x\,}
\newcommand{\inty}{\int d^2y\,}
\newcommand{\intxy}{\int d^2x\,d^2y\,}

\newcommand{\bP}{\bar{\Psi}}
\newcommand{\bc}{\bar{\chi}}
\newcommand{\hs}{\hspace*{0.6cm}}

\newcommand{\bra}{\left\langle}
\newcommand{\ket}{\right\rangle}
\newcommand{\bracket}{\left\langle\,\right\rangle}

\newcommand{\D}{\mbox{$\mathcal{D}$}}
\newcommand{\N}{\mbox{$\mathcal{N}$}}
\newcommand{\Lag}{\mbox{$\mathcal{L}$}}
\newcommand{\V}{\mbox{$\mathcal{V}$}}
\newcommand{\Z}{\mbox{$\mathcal{Z}$}}

\begin{titlepage}


\vspace{2cm}

\begin{center}

{\Large {\bf Non-Local Thirring model with backward\\[2.5mm] and
umklapp interactions}}

\vspace{1.3cm}

V. I. Fern\'andez$^{a,b}$, C. A. Iucci$^{a,b}$ and C. M.
Na\'on$^{a,b}$\footnote{e-mail: victoria@fisica.unlp.edu.ar, iucci@fisica.unlp.edu.ar,
naon@fisica.unlp.edu.ar}

\vspace{.8cm}

$^a$ {\it Instituto de F\'{\i}sica La Plata, Departamento de
F\'{\i}sica, Facultad de Ciencias Exactas, Universidad Nacional de
La Plata.  CC 67, 1900 La Plata, Argentina.}

\smallskip

$^b$ {\it Consejo Nacional de Investigaciones Cient\'{\i}ficas y
T\'ecnicas, Argentina.}

\vspace{1cm}

\begin{abstract}
We extend a non local and non covariant version of the Thirring model in order to
describe a many-body system with backward and umklapp scattering processes. We express
the vacuum to vacuum functional in terms of a non trivial fermionic determinant. Using
path-integral methods we find a bosonic representation for this determinant which
allows us to obtain an effective action for the collective excitations of the system.
By introducing a non local version of the self-consistent harmonic approximation, we
get an expression for the gap of the charge-density excitations as functional of
arbitrary electron-electron potentials. As an example we also consider the case of a
non contact umklapp interaction.
\end{abstract}
\end{center}

\vspace{1 cm}


\noindent{\it Pacs:} 11.10.Lm, 05.30.Fk

\end{titlepage}

\newpage
\section{Introduction}

\hs In recent years there has been much interest in the study of low-dimensional field
theories. In particular they are useful to describe the behavior of strongly
anisotropic physical systems in condensed matter, such as organic conductors
\cite{organic conductors}, charge transfer salts \cite{salts} and quantum wires
\cite{quantum wires}. Probably the two most widely studied 1d systems are the Hubbard
model \cite{Hubbard} and the so-called "g-ology" model \cite{Solyom}. They are known
to display the Luttinger liquid \cite{Haldane} behavior characterized by spin-charge
separation and by non-universal (interaction dependent) power-law correlation
functions.  In a recent series of papers \cite{NLT}, \cite{spinflipping} an
alternative, field-theoretical approach was developed to consider this problem. In
these works a non-local and non-covariant version of the Thirring model
\cite{Thirring} was introduced, in which the fermionic densities and currents are
coupled through bilocal, distance-dependent potentials. This non-local Thirring model
(NLT) contains the Tomonaga-Luttinger (TL) model \cite{TL} as a particular case.
Although it constitutes an elegant framework to analyze the 1d many-body problem, one
serious limitation appears if one tries to make contact with real systems. Indeed, one
has to recall that the building blocks of the NLT are the forward-scattering (fs)
processes which are supposed to dominate the scene only in the low transferred
momentum limit. This means that in its present form it can only provide a very crude
description of the Luttinger liquid equilibrium and transport properties. In ref.
\cite{Fernandez} we started to develop an improved version of the NLT in which larger
momentum transfers are taken into account by including back-scattering (bs)
\cite{back} and umklapp (us) \cite{um}. This formulation led us to a highly
non-trivial action whose physical content was very hard to extract. In the present
paper we reconsider this problem from a different point of view which allowed us to
get insight on the spectrum of the collective modes. By using a non local version of
the self-consistent harmonic approximation (SCHA) \cite{scha} we were able to obtain
general expressions for the bosonic gaps as functionals of arbitrary forward, backward
and umklapp scattering coupling functions. In Section 2 we present the model and
review the steps that allow us to write a purely bosonic action describing the
dynamics of collective excitations. In Section 3 we analyze the bosonic action by
using a self-consistent harmonic approximation. We derive our main formal result, i.e.
the general formula for the gap, valid for pure backscattering ($g_{1}\neq 0 , g_{3} =
0$) and pure umklapp scattering ($g_{3}\neq 0 , g_{1} = 0$). In Section 4, as an
example, we study in some detail the case of a non local umklapp interaction. In
Section 5 we present our conclusions. In the Appendix we review the main points of the
non local SCHA.

\section{The Model}
\setcounter{equation}{0}

\hs In this section we begin the study of an extended version of the NLT which
includes the contribution of forward, backward and umklapp scattering. Following the
formulation proposed in \cite{NLT} we shall attempt to describe these interactions by
means of a fermionic (1+1)-dimensional Quantum Field Theory with Euclidean action
given by

\begin{equation}
S = S_0 + S_{fs} + S_{bs} + S_{us}
\label{1}
\end{equation}
where

\begin{equation}\label{2}
S_0 = \intx\bP i\slp\Psi
\end{equation}
is the unperturbed action associated to a linearized free
dispersion relation. The contributions of the different scattering
processes will be written as

\begin{equation}
S_{fs} = - \frac{g^2}{2} \intxy ( \bP \gamma_{\mu} \Psi ) (x)~
V_{(\mu)}(x,y) ~( \bP \gamma_{\mu} \Psi ) (y) \label{3}
\end{equation}
and

\begin{equation}
S_{bs} + S_{us}= -\frac{{g'}^2}{2} \intxy (\bP~\Gamma_{\mu}
\Psi)(x)\,U_{(\mu)}(x,y)(\bP~\Gamma_{\mu} \Psi)(y) \label{4}
\end{equation}
where the $\gamma_{\mu}'s$ are the usual two-dimensional Dirac matrices and
$\Gamma_{0}=1$, $\Gamma_{1}=\gamma_{5}$. Please keep in mind that no sum over repeated
indices is implied when a subindex $(\mu)$ is involved. Let us also mention that the
coupling potentials $V_{(\mu)}$ and $U_{(\mu)}$ are assumed to depend on the distance
$\mid x-y \mid$ and can be expressed in terms of Solyom's "g-ology" \cite{Solyom} as

\begin{align}
V_{(0)}(x,y) &= \frac{1}{g^2}(g_2+g_4)(x,y) \\ V_{(1)}(x,y) &=
\frac{1}{g^2}(g_2-g_4)(x,y) \\ U_{(0)}(x,y) &= \frac{1}{{g'}^2}(g_3+g_1)(x,y) \\
U_{(1)}(x,y) &= \frac{1}{{g'}^2}(g_3-g_1)(x,y)
\end{align}
In the above equations $g$ and ${g'}$ are just numerical constants that could be set
equal to one. We keep them to facilitate comparison of our results with those
corresponding to the usual Thirring model. Indeed, this case is obtained by choosing
${g'}=0$ and $V_{(0)}(x,y)=V_{(1)}(x,y)=\delta^2(x-y)$. On the other hand, the
non-covariant limit ${g'}=0$, $V_{(1)}(x,y)=0$ gives one version ($(g_2 =g_4$) of the
TL model \cite{TL}.

The terms in the action containing $g_2$ and $g_4$ represent forward scattering
events, in which the associated momentum transfer is small. In the $g_2$ processes the
two branches (left and right-moving particles) are coupled, whereas in the $g_4$
processes all four participating electrons belong to the same branch. On the other
hand, $g_1$ and $g_3$ are related to scattering diagrams with larger momentum
transfers of the order of $2 k_F$ (bs) and $4 k_F$ (us) respectively (this last
contribution is important only if the band is half-filled). For simplicity, throughout
this paper we will consider spinless electrons. The extension of our results to the
spin-$1/2$ case with spin-flipping interactions, though not trivial, could be done by
following the lines of ref. \cite{spinflipping}.

Let us now turn to the treatment of the partition function. At this point we recall
that in ref. \cite{NLT} we wrote the fs piece of the action in a localized way :
\begin{equation} S_{fs} = -\frac{g^2}{2} \intx J_{\mu}  K_{\mu}. \label{7}
\end{equation}

\noindent where $J_{\mu}$ is the usual fermionic current, and $
K_{\mu}$ is a new current defined as

\begin{equation}
 K_{\mu}(x) = \inty V_{(\mu)}(x,y)  J_{\mu}(y).
\label{8} \end{equation} Using a functional delta and introducing auxiliary bosonic
fields in the path-integral representation of the partition function $\Z$, we were
able to write (see \cite{NLT} for details):
\begin{equation} \Z = \N \int
\mbox{$\mathcal{D}$}\bP \mbox{$\mathcal{D}$}\Psi
\mbox{$\mathcal{D}$}\tilde{A}_{\mu}\mbox{$\mathcal{D}$}\tilde{B}_{\mu}~
\exp\left\{-\intx [\bP i\slp\Psi +\tilde{A}_{\mu}\tilde{B}_{\mu} +
\frac{g}{\sqrt{2}}(\tilde{A}_{\mu} J_{\mu} + \tilde{B}_{\mu}K_{\mu})] \right\}
\label{9} \end{equation}

\noindent If we define \begin{equation} \bar{B}_{\mu}(x) = \inty
V_{(\mu)}(y,x)\tilde{B}_{\mu}(y), \label{10} \end{equation}
\begin{equation} \tilde{B}_{\mu}(x) = \inty V^{-1}_{(\mu)}(y,x)
\bar{B}_{\mu}(y), \label{11} \end{equation}

\noindent with $V^{-1}_{(\mu)}(y,x)$ satisfying

\begin{equation} \inty V^{-1}_{(\mu)}(y,x) V_{(\mu)}(z,y) = \delta^2 (x-z),
\label{12} \end{equation}

\noindent and change auxiliary variables in the form

\begin{equation}
A_{\mu}=\frac{1}{\sqrt{2}}(\tilde{A}_{\mu} +\bar{B}_{\mu}),
\label{13}
\end{equation}

\begin{equation}
B_{\mu}=\frac{1}{\sqrt{2}}(\tilde{A}_{\mu} - \bar{B}_{\mu}),
\label{14}
\end{equation}

\noindent we obtain
\begin{equation}
\Z = \N \int \mbox{$\mathcal{D}$}\bP \mbox{$\mathcal{D}$}\Psi
\mbox{$\mathcal{D}$}{A}_{\mu}\mbox{$\mathcal{D}$}{B}_{\mu}~e^{-S(A,B)-S_{bs}-S_{us}}~
\exp\left[-\intx \bP (i\slp- g \lnA)\Psi\right] \label{15}
\end{equation}

\noindent where \beqn S(A,B)=\frac{1}{2}\intxy
V^{-1}_{(\mu)}(x,y)[A_{\mu}(x)
      A_{\mu}(y)-B_{\mu}(x)B_{\mu}(y)]
\label{16}
\eeqn

\noindent The Jacobian associated with the change $(\tilde{A}, \tilde {B})\rightarrow
(A,B)$ is field- independent and can then be absorbed in the normalization constant
$\N$. Moreover, we see that the $B$-field is completely decoupled from both the
$A$-field and the fermion field. Keeping this in mind, it is instructive to try to
recover the partition function corresponding to the usual covariant Thirring model
($V^{-1}_{(0)}(y,x)$ =$ V^{-1}_{(1)}(x,y)$ =$ \delta^2(x-y)$, and ${g'}=0$), starting
from (\ref{15}). In doing so one readily discovers that $B_{\mu}$ describes a
negative-metric state whose contribution must be factorized and absorbed in $\N$ in
order to get a sensible answer for \Z. This procedure parallels, in the path-integral
framework, the operator approach of Klaiber \cite{Thirring}, which precludes the use
of an indefinite-metric Hilbert space. Consequently, from now on we shall only
consider the $A$ contribution.

At this stage we see that, when bs and us processes are disregarded, the procedure we
have just sketched allows us to express $\Z$ in terms of a fermionic determinant. Now
we will show that this goal can also be achieved when the larger momentum transfers
are taken into account. To this end we write:

\begin{equation} S_{bs} + S_{us} = -\frac{{g'}^2}{2} \intx  L_{\mu}  M_{\mu}
\label{17} \end{equation}

\noindent where $L_{\mu}$ and $M_{\mu}$ are fermionic bilinears defined as

\begin{equation}
L_{\mu}(x) = \bP(x)  \Gamma_{\mu} \Psi(x),
\label{18}
\end{equation}

\begin{equation}
 M_{\mu}(x) = \inty U_{(\mu)}(x,y)  L_{\mu}(y).
\label{19}
\end{equation}

\noindent Thus it is evident that we can follow the same
prescriptions as above, with $L_{\mu}$ and $M_{\mu}$ playing the
same roles as $J_{\mu}$ and $K_{\mu}$, respectively. After the
elimination of a new negative metric state whose decoupled
partition function is absorbed in the normalization factor, as
before, one obtains
\begin{equation}
\Z = \N \int  \mbox{$\mathcal{D}$} A_{\mu} \mbox{$\mathcal{D}$} C_{\mu}~ \det (i \slp
- g \lnA - {g'} \Gamma _{\mu}C_{\mu}) ~ \exp\left(-S[A] - S[C]\right) \label{20}
\end{equation}

\noindent where

\beqn S[A_{\mu}] = \frac{1}{2}\intxy A_{\mu}(x) V^{-1}_{(\mu)}A_{\mu}(y) \nonumber \\
S[C_{\mu}] = \frac{1}{2}\intxy C_{\mu}(x) U^{-1}_{(\mu)}C_{\mu}(y) \nonumber
\\ \label{21} \eeqn

\noindent and \begin{equation} \inty U^{-1}_{(\mu)}(y,x) U_{(\mu)}(z,y) = \delta^2
(x-z), \label{22} \end{equation}

Then we have been able to express $\Z$ in terms of a fermionic determinant. Let us
stress, however, that this determinant is a highly non trivial one. Indeed, the term
in ${g'}$ is not only a massive-like term (in the sense that it is diagonal in the
Dirac matrices space) but it also depends on the auxiliary field $C_{\mu}(x)$. As
shown in \cite{Fernandez} one can combine a chiral change in the fermionic
path-integral measure with a formal expansion in ${g'}$ in order to get a bosonic
representation for the fermionic determinant. Let us start by performing the following
transformation:

\begin{align}
\Psi(x) =& e^{g[\gamma_5 \Phi(x) - i \eta(x)]} ~ \chi(x)  \\
\bP(x) =& \bar\chi(x) ~ e^{g[\gamma_5 \Phi(x) + i\eta(x)]}
\label{23}
\end{align}

\begin{equation}
\mbox{$\mathcal{D}$} \bP   \mbox{$\mathcal{D}$} \Psi = J
[\Phi,\eta]  \mbox{$\mathcal{D}$} \bar\chi
\mbox{$\mathcal{D}$}\chi, \label{24}
\end{equation}
where $\Phi$ and $\eta$ are scalar fields and $J[\Phi,\eta]$ is
the Jacobian of the transformation. As it is well known, the above
transformation permits to decouple the field $A_{\mu}$ from the
fermionic fields if one writes

\begin{equation}
A_{\mu}(x) = \partial_{\mu}\eta(x) + \epsilon_{\mu
\nu}\partial_{\nu}\Phi(x) \label{25}
\end{equation}
which can also be considered as a bosonic change of variables with
trivial (field independent) Jacobian. As a result we find

\begin{equation}
\det (i \slp - g \lnA -{g'} \Gamma _{\mu}C_{\mu}) = J[\Phi,\eta] \det (i\slp-{g'}
e^{2g \gamma_5 \Phi} \Gamma _{\mu}C_{\mu}) \label{26}
\end{equation}
After a suitable regularization the fermionic Jacobian reads \cite{Chen-Lee}

\begin{equation}
J[\Phi,\eta]= \exp{\frac{g^2}{2 \pi}\intp\left[-p_{1}^2 \Phi (p)
\Phi (-p) + p_{1}^2 \eta (p) \eta (-p)- 2 p_{0} p_{1} \Phi (p)
\eta (-p)\right]}. \label{27}
\end{equation}
The vacuum to vacuum functional is then expressed as

\begin{equation}\label{28}
\Z= \N'\int  \mbox{$\mathcal{D}$}\Phi \mbox{$\mathcal{D}$}\eta
\mbox{$\mathcal{D}$}C_{\mu} e^{-(S[\Phi,\eta]+S[C_{\mu}])} J[\Phi,\eta] \det
(i\slp-{g'} e^{2g \gamma_5 \Phi} \Gamma _{\mu}C_{\mu})
\end{equation}
where $S[\Phi,\eta]$ arises when one inserts (\ref{25}) in
$S[A_{\mu}]$ (See equation (\ref{21})):

\begin{equation}
S[\Phi, \eta]= \intp[ \Phi (p) \Phi (-p) A(p) + \eta (p) \eta (-p)
B(p) + \Phi (p) \eta (-p) C(p) ]
\end{equation}
with

\begin{align}
A(p)&= \frac{1}{2}\left[p_0 ^2 \hat{V}_{(1)}^{-1} (p) + p_1 ^2 (\hat{V}_{(0)}^{-1} (p)
+ \frac{g^2}{\pi})\right] \\ B(p)&=\frac{1}{2}\left[p_0 ^2 \hat{V}_{(0)}^{-1} (p) +
p_1 ^2 (\hat{V}_{(1)}^{-1} (p) - \frac{g^2}{\pi})\right] \\
C(p)&=p_0p_1\left(\hat{V}_{(0)}^{-1} (p)- \hat{V}_{(1)}^{-1} (p) +
\frac{g^2}{\pi}\right).
\end{align}

 \hs The fermionic determinant in the
above expression can be analyzed in terms of a formal perturbative expansion. Indeed,
taking ${g'}$ as perturbative parameter, and using the fermionic fields $\chi$ and
$\bar\chi$ defined in (\ref{23}) one can write

\begin{equation} \Z_F=
\sum_{n=0}^{\infty}\frac{{g'}^{n}}{n!}\bra\prod_{j=1}^{n} \int d^2x_j\, \bc
(x_{j})~\mbox{$ \mathsf{C} $} (x_{j}) ~\chi (x_{j}) \ket_0 \label{29} \end{equation}

where, for later convenience we have defined \beqn \Z_F&=& \det (i\slp-{g'} e^{2g
\gamma_5 \Phi} \Gamma _{\mu}C_{\mu}) \nonumber\\ \label{30} \eeqn \noindent and \beqn
\mathsf{C}= \left(
\begin{array}{cc}
C_{+} &  0 \\
0 & C_{-}
\end{array} \right)
\label{31} \eeqn \noindent with \beqn
 \left \{  \begin{array}{l}
C_{+}= (C_0 + C_1) ~ e^{2g \Phi (x)} \\
C_{-}= (C_0 - C_1) ~ e^{-2g \Phi (x)}
 \end{array} \right.
\label{32}
\eeqn

\hs By carefully analyzing each term in the series we found a
selection rule quite similar to the one obtained in the
path-integral treatment of (1+1) massive fermions with local
\cite{Coleman} \cite{Naon} and non-local interactions
\cite{Li-Naon}. Indeed, due to the fact that in (\ref{29})
$\bra~\ket_0$ means v.e.v. with respect to free massless fermions,
the v.e.v.'s corresponding to $j = 2k+1$ are zero. Thus, we obtain

\begin{equation}
\begin{split}
\Z_F=&\sum_{k=0}^{\infty}\frac{({g'}c \rho)^{2k}}{(k!)^2 (2\pi)^{2k}} \int
\prod_{i=1}^{k} d^2x_i\, d^2y_i\\ \times & \prod_{i=1}^{k} \left[C_0 ( x_{i}) + C_1 (
x_{i})\right]\left[C_0 ( y_{i}) - C_1 ( y_{i})\right] \\ \times & \exp{2g
\sum_{i=1}^{k} \left[ \Phi (x_i) - \Phi ( y_i)\right]} \\ \times & \frac{
\prod_{i>j}^{k}(c \rho)^4 | x_{i}- x_{j}|^2 | y_{i}- y_{j}|^2 } {\prod_{i,j}^{k}(c
\rho)^2|x_{i}- y_{j}|^2} \label{33}
\end{split}
\end{equation}
where $\rho$ is a normal ordering parameter and $c$ is related to
Euler's constant.

In order to obtain a bosonic description of the present problem we shall now propose the
following bosonic Lagrangian density:

\begin{equation}
\mbox{ $\mathcal{L}$ }_B= \frac{1}{2} (\partial _{\mu} \varphi )^2 +
\frac{\alpha_0}{2\beta^2} (m_{+} e^{i\beta \varphi} + m_{-} e^{ -i \beta
\varphi})
\label{34}
\end{equation}

\noindent with $ \beta$, $m_{+}(x)$ and $m_{-}(x)$ to be
determined. The quantity $\alpha_0$ is just a constant that we
include to facilitate comparison of our procedure with previous
works on local bosonization \cite{Coleman} \cite{Naon}. Please
notice that for $m_{+} = m_{-} = 1$ this model coincides with the
well known sine-Gordon model that can be used to describe a
neutral Coulomb gas. In this context $\frac{\alpha_0}{\beta^2}$ is
nothing but the corresponding fugacity \cite{Samuel}. We shall now
consider the partition function

\begin{equation} \Z_B=  \int
\mbox{$\mathcal{D}$} \varphi ~e^{- \intx  \mbox{ $\mathcal{L}$ }_B} \label{35}
\end{equation}
and perform a formal expansion taking the fugacity as perturbative
parameter. It is quite straightforward to extend the analysis of
each term, already performed for $m_{+} = m_{-} = 1$, to the
present case in which these objects are neither equal nor
necessarily constants. The result is

\begin{multline}
\Z_B=\sum_{l=1}^{\infty}
\frac{1}{(l!)^2}\left(\frac{\alpha_0}{2\beta^2}\right)^{2l} \int
\left(\prod_{i=1}^l d^2x_i\, d^2y_i\right)\left( \prod _{i=1}^l
m_{+}(x_i) ~m_{-}(y_i)\right)\left(\frac{\rho}{\Lambda}\right)^{2l
\frac{\beta^2}{4 \pi} }\\ \times
\frac{\prod_{i>j}^{l}\left[(c\rho)^2|x_{i}-x_{j}|~|y_{i}-y_{j}|\right]^{\beta^2/2\pi}}
{\prod_{i,j}^{l}\left[(c\rho)|x_{i}-y_{j}|\right]^{\beta^2/2\pi}}.
\label{36}
\end{multline}

Comparing  this result with equation (\ref{33}), we see that both series coincide if
the following identities hold: \beqn \beta &=& \pm 2\sqrt{\pi} \nonumber\\
\frac{\alpha_0}{\beta^2}&=& \frac{{g'}\Lambda c}{\pi} \label{37} \eeqn \noindent and

\beqn
 m_{+}(x_i)&=& \left(C_0(x_i ) + C_1(x_i )\right) ~ e^{2g \Phi (x_i )} \nonumber\\
 m_{-}(y_i )&=& \left(C_0 (y_i )- C_1 (y_i )\right) ~ e^{-2g \Phi (y_i )}
\label{38} \eeqn

Thus we have found a bosonic representation for the fermionic determinant (\ref{33}).
This is given by (\ref{35}) together with the identities (\ref{37}) and (\ref{38}).
Let us emphasize that equations (\ref{37}) are completely analogous to the
bosonization formulae first obtained by Coleman \cite{Coleman} whereas equations
(\ref{38}) constitute a new result, specially connected to the present problem.

\section{Derivation of gap equations through self consistent harmonic approximation}
\setcounter{equation}{0}

Inserting (\ref{35}) in (\ref{28}) we can express the partition
function of this system in terms of five scalars: $\Phi$, $\eta$,
$C_0$, $C_1$ and $\varphi$. We are thus led to a completely
bosonized action $S_{bos}$:

\begin{multline}
S_{bos}=\intp [ \Phi (p) \Phi (-p) A(p) + \eta (p) \eta (-p) B(p) + \Phi (p) \eta
(-p)C(p) +\varphi (p)\varphi (-p)\frac{p^2}{2}]\\ +\frac{1}{2}\intxy C_{\mu} (x)
U^{-1}_{(\mu)}(x,y) C_{\mu} (y) +\frac{{g'} \Lambda c}{\pi} \intx  [ C_{(0)}
(x)f_{0}(x) + i C_{(1)} (x)f_{1}(x)] \label{39}
\end{multline}
where

\begin{align}
f_{0}(x)=&\cos ((\sqrt{4 \pi}\varphi -2i g \Phi )(x))\\
f_{1}(x)=& \sin (( \sqrt{4 \pi}\varphi - 2i g \Phi )(x))
\label{40}
\end{align}

Since the integrals in $C_0$ and $C_1$ are quadratic these fields are easily
integrated out and one gets

\begin{equation}
\Z = \N \int \mbox{$\mathcal{D}$} \Phi ~ \mbox{$\mathcal{D}$} \eta
~ \mbox{$\mathcal{D}$} \varphi ~ e^{-S_{eff}[\Phi,\eta,\varphi]}
\label{41}
\end{equation}
with

\begin{equation}
S_{eff}[\Phi,\eta,\varphi ] = S_{0} + S_{int}
\label{42}
\end{equation}
where

\begin{equation}
S_{0}=\intp [ \Phi (p) \Phi (-p) A(p) + \eta (p) \eta (-p) B(p)
 +  \Phi (p) \eta (-p) C(p) + \varphi (p)\varphi
(-p)\frac{p^2}{2}]
\end{equation}

\begin{multline}\label{43}
S_{int}=-\frac{(\Lambda c)^2}{2\pi^2}\intxy g_{1} (x,y) \cos\left[
\sqrt{4\pi}(\varphi(x) -\varphi (y)) -2i g (\Phi(x) -\Phi(y))\right]\\
-\frac{(\Lambda c)^2}{2\pi^2}\intxy g_{3} (x,y) \cos\left[\sqrt{4
\pi}(\varphi(x) + \varphi (y)) -2i g (\Phi(x) + \Phi(y))\right].
\end{multline}

It is now convenient to  diagonalize the quadratic part of the effective action by
introducing the fields $\zeta$, $\chi$ and $\xi$:

\begin{align}\label{eq:changeVariables}
\Phi&=\frac{i\zeta}{\tilde{g}}+\frac{2i\tilde{g}Bp^2}{\Delta+2B\tilde{g}^2p^2}\xi\\
\eta&=\frac{-iC}{2B\tilde{g}}\zeta -\frac{i\tilde{g}Cp^2}{\Delta +
2B\tilde{g}^2p^2}\xi +\frac{1}{\tilde{g}}\chi \\
\varphi&= -
\zeta+\frac{\Delta}{\Delta+2B\tilde{g}^2p^2}\xi,
\end{align}
where we have defined $\tilde{g}^2=g^2/\pi$ and $\Delta(p)=C(p)^2-4A(p)B(p)$. We then
obtain

\begin{equation}
S_{0}=\frac{1}{2}\intp\left[\zeta(p)\left(p^2+\frac{\Delta}{2B\tilde{g}^2}\right)\zeta(-p)
+\chi(p)\frac{2B}{\tilde{g}^2}\chi(-p)+
\xi(p)\frac{p^2\Delta}{\Delta + 2B\tilde{g}^2p^2}\xi(-p)\right],
\label{44}
\end{equation}

\begin{multline}
S_{int}=-\frac{(\Lambda c)^2}{2\pi^2}\intxy g_{1} (x,y)
\cos\sqrt{4\pi}[\xi(x) -\xi (y)]\\-\frac{(\Lambda
c)^2}{2\pi^2}\intxy g_{3} (x,y) \cos\sqrt{4\pi}[\xi(x) + \xi
(y)].\label{45}
\end{multline}

One can see that the $\zeta$ and $\chi$ fields become completely
decoupled from $\xi$. Moreover, it becomes apparent that the
$\xi$-dependent piece of the action $S_{int}$ is the only one
containing relevant contributions (i.e. gapped modes):

\begin{multline}
S[\xi]=\intp\xi(p)\frac{F(p)}{2}\xi(-p)
-\frac{(\Lambda c)^2}{2\pi^2}\intxy g_1(x,y)\cos\sqrt{4\pi}[\xi(x)-\xi(y)]\\
-\frac{(\Lambda c)^2}{2\pi^2}\intxy
g_3(x,y)\cos\sqrt{4\pi}[\xi(x)+\xi(y)] \label{46}
\end{multline}
with

\begin{align}
F(p)=&\frac{1}{Kv}(p_0^2+v^2p_1^2)\\ K=&\sqrt{\frac{1+ g_{4}/ \pi -g_{2}/ \pi}{1+
g_{4}/ \pi + g_{2}/ \pi}}\\
v=&\sqrt{(1+\frac{g^2\hat{V}_{(0)}(p)}{\pi})(1-\frac{g^2\hat{V}_{(1)}(p)}{\pi})}.\label{47}
\end{align}

Unfortunately, even if we succeeded in simplifying the original action, $S[\xi]$ is
not yet soluble. We introduce a non-local version of the self-consistent harmonic
approximation \cite{scha}. Basically, this amounts to replacing the so called {\it{
true}} action (\ref{46}) by a {\it{trial}} action in which the cosine terms are
approximated as

\begin{equation}
-\frac{(\Lambda c)^2}{2\pi^2} g_{1,3} (x,y)
\cos\sqrt{4\pi}[\xi(x)\pm\xi(y)] \longrightarrow
\frac{\Omega_{1,3}(x,y)}{2}\,\xi(x)\xi(y) \label{48}
\end{equation}
where the functions $\Omega _{i}$ of the trial action can be
variationally determined (See the Appendix for details). We will
consider two cases: pure backscattering ($g_{1}\neq 0 , g_{3} =
0$) and pure umklapp scattering ($g_{3}\neq 0 , g_{1} = 0$). Once
this is done, it is straightforward to obtain the charge spectrum.
Indeed, going to momentum space, and back to real frequencies,
$p_{0}= i \omega , p_{1}= k $, the following equations are
obtained:

\begin{equation}
Kv\Omega_i-\omega^2 + v^2 k^2 = 0 .\label{49}
\end{equation}

The gap equations satisfied by $\Omega_{i}$ are:

\begin{equation}
\Omega_{1,3}(p) = \frac{4\Lambda^2 c^2}{\pi}\intx g_{1,3}(x) e^{-4\pi[I_1(x)\mp
I_1(0)]}(e^{ipx}\mp1) \label{50}
\end{equation}
where

\begin{equation}
I_1 (x)= \intp \frac{e^{-ipx}}{F(p)+\Omega_i(p)}\label{51}
\end{equation}

Equation (\ref{50}) is our main formal result. It gives, within
the SCHA, a closed expression for the gap as functional of any
arbitrary strength of the interactions (the $ g_{i}'s$). Let us
study, as examples, the cases where umklapp and backscattering are
local interactions (i.e. $ g_{^{1}_{3}}(x) = g_{^{1}_{3}}\delta ^2
(x)$), but keeping arbitrary fs interactions. It is easy to see
that the pure backscattering process is gapless ($\Omega_{1}=0$)
and that the umklapp interaction opens a gap in the charge density
spectrum ($\Omega _{3}= \text{constant}$), as expected \cite{back}
\cite{um}. For this last case we obtain

\begin{equation} \Omega_{3}=\frac{8\Lambda^2c^2}{\pi}g_3
e^{-8\pi I_1(0)} \label{52}
\end{equation}
As $\Omega_3$ is a constant, i.e. it does not depend on $p$ we can calculate the
propagator in terms of $\Omega_3$, and the equation then reduces to an algebraic
equation for $\Omega_3$. If we consider also contact fs potentials (i.e. $ g_{2,4}(x)
= g_{2,4}\delta ^2(x)$), the integral $I_1(0)$ can be explicitly performed. After a
suitable regularization one obtains:

\begin{equation}\Omega_3=
\frac{4\Lambda^2}{K}\left(\frac{2c^2K}{\pi}g_3\right)^{\frac{1}{1-2K}}.
 \label{53}
\end{equation}

An equation completely analogous to (\ref{52}) was found in
\cite{spinflipping} by working directly with a local coupling
associated to spin-flipping interactions in that case. This
provides a consistency check for our computation. Of course,
having derived a general formula for the gap, like (\ref{50}), it
is interesting to analyze the effect of a non-contact interaction
on the gap. We present this study in the next section.

\section{Gap of charge-density modes for a non-local umklapp coupling}
\setcounter{equation}{0}

The purpose of this Section is to find a solution to equation (\ref{50}) for a pure
umklapp potential of the form
\begin{equation}
g_3(p)=\lambda_0+\frac{\epsilon}{\Lambda^2} p_1^2.
\end{equation}
being $\epsilon\ll 1$. It consists of the local potential plus a small non local
correction. We start by writing the potential in coordinate space:

\begin{equation}
g_3(x)=\lambda_0\delta^2(x)-\frac{\epsilon}{\Lambda^2}\partial_1^2\delta^2(x)
\end{equation}
where $\partial_1^2\delta^2(x)$ is the second derivative of the delta function with
respect to $x_1$ only. By replacing this function in (\ref{50}) and integrating we
obtain

\begin{equation}
\Omega(p)= \frac{4 c^2\Lambda^2}{\pi}\left[2\lambda_0 e^{-8\pi
I_1(0)}-\frac{\epsilon}{\Lambda^2}
\partial_1^2\left(e^{-4\pi\left[I_1(0)+I_1(x)\right]}
(1+e^{ipx})\right)_{x=0}\right]
\end{equation}
where we have dropped the subindex 3 in $\Omega(p)$. Performing the derivation and
rearranging terms we can write

\begin{equation}\label{Omega}
\Omega(p)=\frac{4\Lambda^2c^2}{\pi}\left[2\lambda_0+8\pi\frac{\epsilon}{\Lambda^2}
\partial_1^2I_1(0)\right]e^{-8\pi I_1(0)}+
\frac{4c^2\epsilon}{\pi}e^{-8\pi I_1(0)}p_1^2
\end{equation}
In this last equation we have disregarded terms proportional to $\partial_1I_{1}(0)$
which is zero by symmetry. We observe that $\Omega$ retains the same structure as
$g_3(p)$ i.e. it is of the form

\begin{equation}\label{OmegaForm}
\Omega(p)=\mu^2+gp_1^2
\end{equation}
where $\mu^2$ and $g$ are two parameters to be determined in terms of $\lambda_0$ and
$\epsilon$. The fact that $\Omega$ preserves its simple form allows to calculate the
propagator, as occurs in the local case, and in consequence, to transform the equation
into an algebraic one, easier to deal with. The objects needed are the propagator in
the origin $I_{1}(0)$ and its second derivative in the origin $\partial_1^2I_{1}(0)$.
Theirs expression are

\begin{equation}\label{proporig}
I_{1}(0)=\frac{-Kv}{4\pi\sqrt{v^2+Kgv}}\ln\left[\frac{\mu^2Kv}{4\Lambda^2(v^2+Kgv)}\right],
\end{equation}
and

\begin{equation}\label{2proporig}
\partial_1^2I_{1}(0)=\frac{-\Lambda^2Kv}{8\pi\sqrt{v^2+Kgv}}
\left\{\frac{\mu^2Kv}{\Lambda^2(v^2+Kgv)}
\ln\left[\frac{e\mu^2Kv}{4\Lambda^2(v^2+Kgv)}\right]-2\right\}.
\end{equation}
Replacing (\ref{OmegaForm}), (\ref{proporig}) and (\ref{2proporig}) in equation
(\ref{Omega}), and equating the coefficients of each power of $p$, we obtain the pair
of equations

\begin{equation}\label{eq1}
x=\frac{gKv}{2\epsilon\sqrt{v^2+Kgv}}\left[\lambda_0-\frac{\epsilon
Kv}{\sqrt{v^2+Kgv}}(2x\ln ex+1)\right]
\end{equation}

\begin{equation}\label{eq2}
g=\frac{4\epsilon c^2}{\pi}x^{2Kv/\sqrt{v^2+Kgv}}
\end{equation}
where we have introduced the adimensional quantity
$x=\mu^2Kv/4\Lambda^2(v^2+Kgv)$. These coupled equations are to be
solved for $g$ and $x$.

Of course, a complete solution of (\ref{eq1}) and (\ref{eq2}) is very difficult to
obtain, but we can obtain an approximate solution in the limit $\epsilon\ll 1$. It
corresponds to take a small non local correction to the local potential. For
$\epsilon=0$ the solution is $g=0$, and we expect no discontinuities as $\epsilon$
grows from that value. So for small $\epsilon$ the solution for $g$ is also small.
This allows to approximate $v^2+Kgv\approx v^2$, and one then gets

\begin{equation}
x=\frac{2c^2K}{\pi}x^{2K}\left[\lambda_0-\epsilon K(2x\ln
ex+1)\right]
\end{equation}

\begin{equation}
g=\frac{4c^2\epsilon}{\pi}x^{2K}
\end{equation}
We notice that the first equation does not depend on $g$, though it is still difficult
to solve. We shall employ a perturbative approach. Suppose $x_0$ is the solution to
the $\epsilon=0$ equation:

\begin{equation}
x_0=\left(\frac{2c^2 K \lambda_0}{\pi}\right)^{\frac{1}{1-2K}}
\end{equation}
We can try a solution of the form $x=x_0+\delta x$ where $\delta x$ goes to zero as
$\epsilon$ approaches zero. Replacing this form into the equation, and retaining terms
up to first order in $\delta x$ we obtain

\begin{equation}
\delta x=\frac{-2\epsilon x_0(2x_0\ln ex_0+1)}{\lambda_0(1-2K)}
\end{equation}

\begin{equation}
g=\frac{2\epsilon x_0}{K\lambda_0}
\end{equation}

The first equation gives the correction to the gap of charge-density modes due to
non-contact umklapp interactions. If one introduces a small non local perturbation, it
is enough to modify the value of the gap. Note that the numerator can be positive or
negative, depending on the value of $x_0$, so the same happens with the modification
in the gap. It can be lowered or increased by umklapp non local interactions. The
second equation gives the correction to the kinetic part of the spectrum.

\section{Final result and next steps}
\setcounter{equation}{0}

In this paper we have presented a non local and non covariant extension of the
Thirring model which can be used to describe a 1d many-body system when not only
forward but also backward and umklapp scattering is considered. In this sense our
results improve previous formulations reported in \cite{NLT} and \cite{Fernandez}. We
were able to write the vacuum to vacuum functional of this model in terms of a
non-trivial fermionic determinant. We obtained a bosonic representation for this
determinant which led us to an effective action associated to the elementary
collective excitations of the system. By employing the self-consistent harmonic
approximation we found a closed expression for the gap of the charge-density spectrum
as functional of arbitrary (not necessarily local) two-body potentials. This equation
is valid for purely backward ($g_{1}\neq 0 , g_{3} = 0$) or pure umklapp scattering
($g_{3}\neq 0 , g_{1} = 0$). As an application of this formula we studied the effect
of a non local umklapp potential of the form $g_3(p)=\lambda_0+ (\epsilon/\Lambda^2)
p_1^2$ for $\epsilon$ small. Besides its illustrative purpose, the result obtained in
this example is interesting in itself because most of the previous investigations
involving umklapp scattering do not consider non local effects \cite{um-news}.

\section*{Acknowledgements}

This work was partially supported by Universidad Nacional de La Plata  and Consejo
Nacional de Investigaciones Cient\'{\i}ficas y T\'ecnicas, CONICET (Argentina).

\section{Appendix: details of the SCHA method}
\setcounter{equation}{0}

We shall give an sketch of the SCHA method. One usually starts
from a partition function

\begin{equation}
\Z\true =\int\D\mu e^{-S\true}
\end{equation}

\noindent where $\D\mu$ is a generic integration measure. An
elementary manipulation leads to

\begin{equation}\label{eq:Z}
\Z\true=\frac{\int\D\mu
e^{-(S\true-S\trial)}\,e^{-S\trial}}{\int\D\mu
e^{-S\trial}}\int\D\mu e^{-S\trial}=\Z\trial\bra
e^{-(S\true-S\trial)} \ket\trial.
\end{equation}

\noindent for any trial action $S\trial$. Now, by means of the
property

\begin{equation}
\bra e^{-f} \ket\geq e^{-\bra f \ket},
\end{equation}

\noindent for $f$ real, and taking natural logarithm in equation
(\ref{eq:Z}), we obtain Feynman's inequality \cite{Feynman}

\begin{equation}\label{eq:Feynman}
\ln\Z \true\geq \ln\Z\trial - \bra S\true-S\trial\ \ket\trial
\end{equation}

We shall study a very general class of true bosonic actions of non
local sine Gordon type, given by

\begin{equation}\label{TrueAction}
S\true=\intp\varphi(p)\frac{F(p)}{2}\varphi(p)- \intxy\frac{\alpha(x-y)}{b^2}\cos b
\frac{\varphi(x)+\varphi(y)}{2},
\end{equation}
where $F(p)$ and $\alpha(x,y)$ are arbitrary functions. The usual local sine Gordon
theory is obtained by taking $F(p)=p^2$ and $\alpha(x,y)=\alpha_0 \delta^{(2)}(x-y)$
whereas (\ref{TrueAction}) reduces to our model (\ref{46}) without backscattering
interactions (i.e. $g_1=0$) by taking

\begin{equation}
F(p)=\frac{1}{Kv}(p_0^2+v^2p_1^2)
\end{equation}

\begin{equation}
b^2=16\pi
\end{equation}

\begin{equation}
\frac{\alpha(x,y)}{b^2}=\frac{(\Lambda c)^2}{2\pi^2}g_3(x,y)
\end{equation}

The trial action we will consider is a quadratic one:

\begin{equation}
S\trial=\intp\varphi(p)\frac{F(p)}{2}\varphi(p)-
\intxy\frac{\Omega(x-y)}{2}\varphi(x)\varphi(y)
\end{equation}
where the function $\Omega$ can be determined by maximizing the
right hand side of equation (\ref{eq:Feynman}). In order to
achieve this goal we first write

\begin{align}
\ln\Z\trial=&\ln\int\D\varphi\exp{\left[-\frac{1}{2}\intx\varphi(x)
(\hat{A}\varphi)(x)\right]} \\=&
\ln(\det\hat{A})^{-1/2}+\text{const} \\=&-\frac{1}{2}\tr\ln\hat{A}
+ \text{const}.
\end{align}
where the operator $\hat{A}$ is defined, in Fourier space, by

\begin{equation}
(\hat{A}\varphi)(p)=[F(p)+ \Omega(p)]\varphi(p).
\end{equation}
being $\Omega(p)$ is the Fourier transform of $\Omega(z)$. It is
then easy to get

\begin{equation}
\tr\ln\hat{A}=\V\intp\ln[F(p)+\Omega(p)]\equiv\V I_0[\Omega]
\end{equation}
where $\V$ is the volume (infinite) of the whole space $\intx$. On
the other hand, it is straightforward to compute $\bra
S\true-S\trial \ket$, by following, for instance, the steps
explained in ref. \cite{Li-Naon}. The result is

\begin{equation}\label{difS}
-\bra S\true-S\trial\ket\trial = \V\intx \left[\frac{\alpha(x)}{b^2} e^{-\uncuarto
b^2\left[I_1(x)+I_1(0)\right]} +\frac{\Omega(x)}{2}I_1(x)\right]
\end{equation}
where $I_1(x)$ is a functional of $\Omega$ given by the propagator of the trial
action:

\begin{equation}
I_1(x)=\intp\,\frac{e^{-ipx}}{F(p)+\Omega(p)}.
\end{equation}
and we have made use of the traslational invariance of $\alpha$ and $\Omega$. Finally
we can gather all the terms, and write them as

\begin{multline}\label{eq:elegant}
\ln\Z\trial - \bra S\true-S\trial\ket\trial = \frac{\V}{2}\intx\frac{\alpha(x)}{b^2}
e^{-\uncuarto b^2\left[I_1(x)+I_1(0)\right]} \\ +
\frac{\V}{2}\intp\frac{\Omega(-p)}{F(p)+\Omega(p)}-\frac{\V}{2}I_0[\Omega] + \const.
\end{multline}
where we have turned the last term of equation (\ref{difS}) into Fourier space. Now
extremizing expression (\ref{eq:elegant}) with respect to $\Omega$ by functional
derivation with respect to $\Omega(k)$, we finally obtain the gap equation

\begin{equation}\label{GapEquation}
\Omega(p)=\frac{1}{2}\intx \alpha(x) e^{-\uncuarto
b^2\left[I_1(x)+I_1(0)\right]}(e^{-ipx}+1)
\end{equation}

If instead of considering only umklapp interactions we consider
only backscattering interactions ($g_3=0$) then we have to change
$\varphi(x)+\varphi(y)\rightarrow\varphi(x)-\varphi(y)$ into the
cosine of equation (\ref{TrueAction}), and the resulting gap
equation is

\begin{equation}\label{GapEquationBack}
\Omega(p)=\frac{1}{2}\intx \alpha(x) e^{-\uncuarto
b^2\left[I_1(x)-I_1(0)\right]}(e^{-ipx}-1)
\end{equation}

\end{document}